\documentclass[twocolumn,showpacs,preprintnumbers,amsmath,amssymb]{revtex4}
\usepackage{graphicx}
\usepackage{amssymb}
\usepackage{dcolumn}
\usepackage{bm}
\usepackage{epstopdf}

\begin{document}

\preprint{APS/123-QED}

\title{ Electronic properties of a graphene antidot in magnetic fields}
\author{P.S. Park$^{1}$, S.C. Kim$^{1}$, and S.-R. Eric Yang$^{1,2}$\footnote{ corresponding author, eyang@venus.korea.ac.kr}}
\affiliation{
$^{1}$Physics Department, Korea  University, Seoul Korea\\
$^{2}$Korea Institute for Advanced Study, Seoul Korea \\
}
\date{\today}

\begin{abstract}
We report on several unusual properties of a graphene antidot created by a piecewise constant
potential in a magnetic field. We find that the total probability of finding the electron in the
barrier can be nearly one while it is almost zero outside the barrier. In addition, for each
electron state of a graphene antidot there is a dot state with exactly the same wavefunction but
with a different energy. This symmetry is a consequence of Klein tunneling of Dirac electrons.
Moreover, in zigzag nanoribbons we find strong coupling between some antidot states and zigzag edge
states. Experimental tests of these effects are proposed.
\end{abstract}


\maketitle

\section{Introduction}

Bulk Landau energy levels  of graphene, $E_n=\pm E_c\sqrt{2|n|}$, have   different magnetic field
$B$ and quantum number $n$ dependence than those of  ordinary two-dimensional Landau levels,
$E_n=\hbar\omega_c(n+1/2)$ (Here  the cyclotron energy $\hbar \omega_c\propto B$,
 where $B$ is the  magnetic field.  The
characteristic energy of Dirac electrons $E_c\propto
\sqrt{B}$). The
wavefunctions of these two systems have, respectively, two-component and one-component  structures\cite{Ando}.
In the presence of a dot or
antidot potential these states of the two systems will be perturbed rather differently.

\begin{figure}[!hbpt]
\begin{center}
\includegraphics[width=0.22\textwidth]{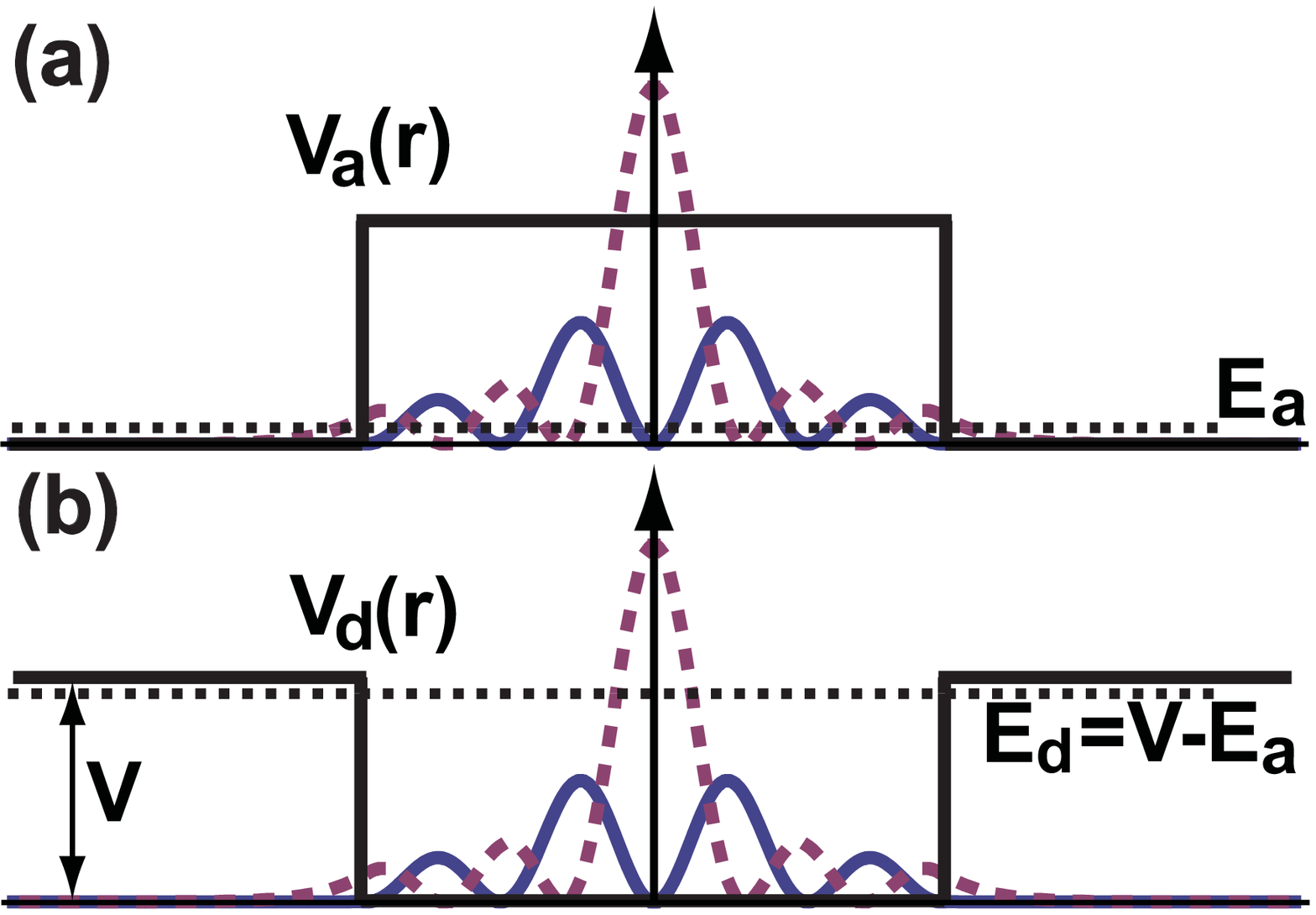}
\includegraphics[width=0.22\textwidth]{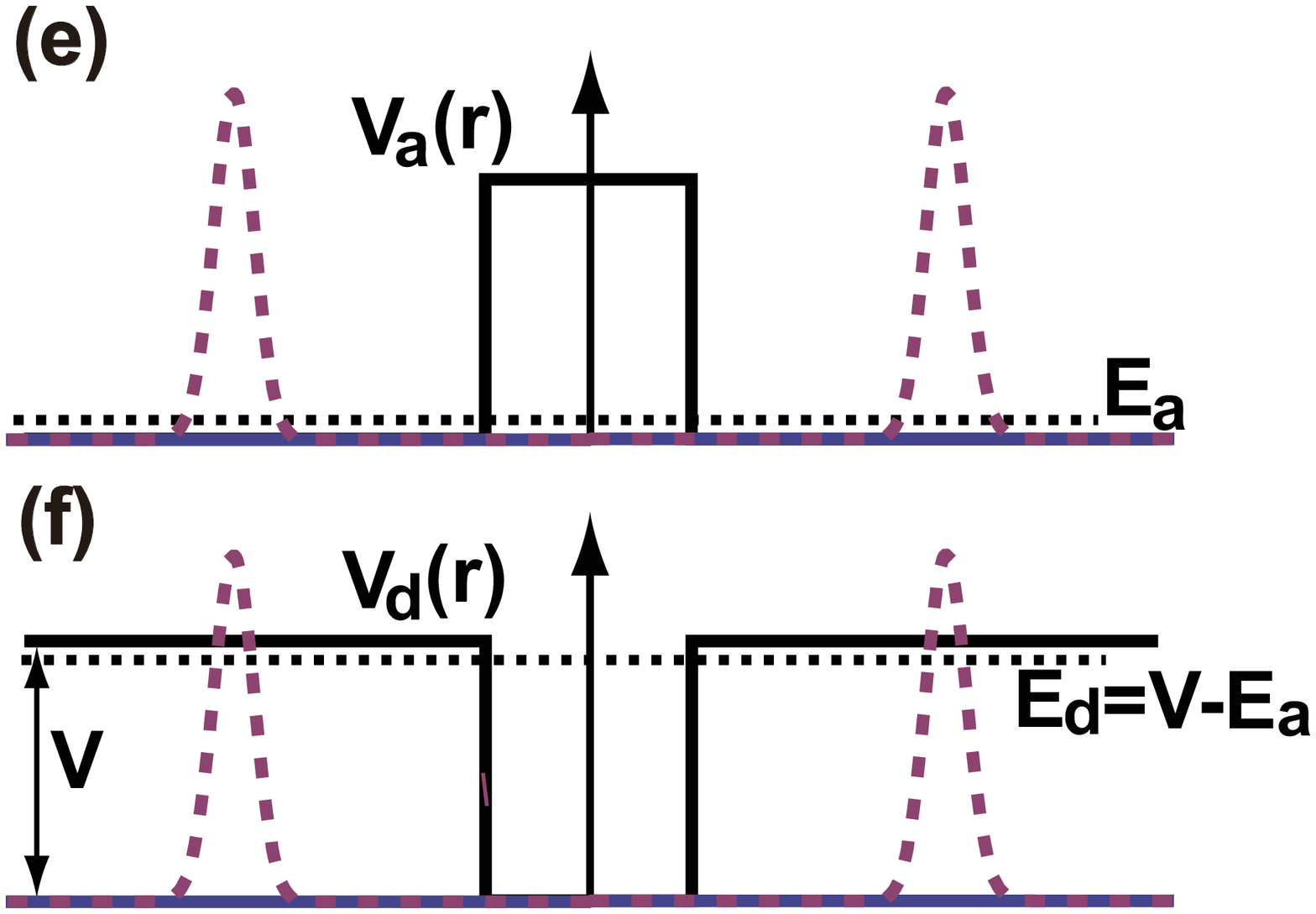}
\includegraphics[width=0.22\textwidth]{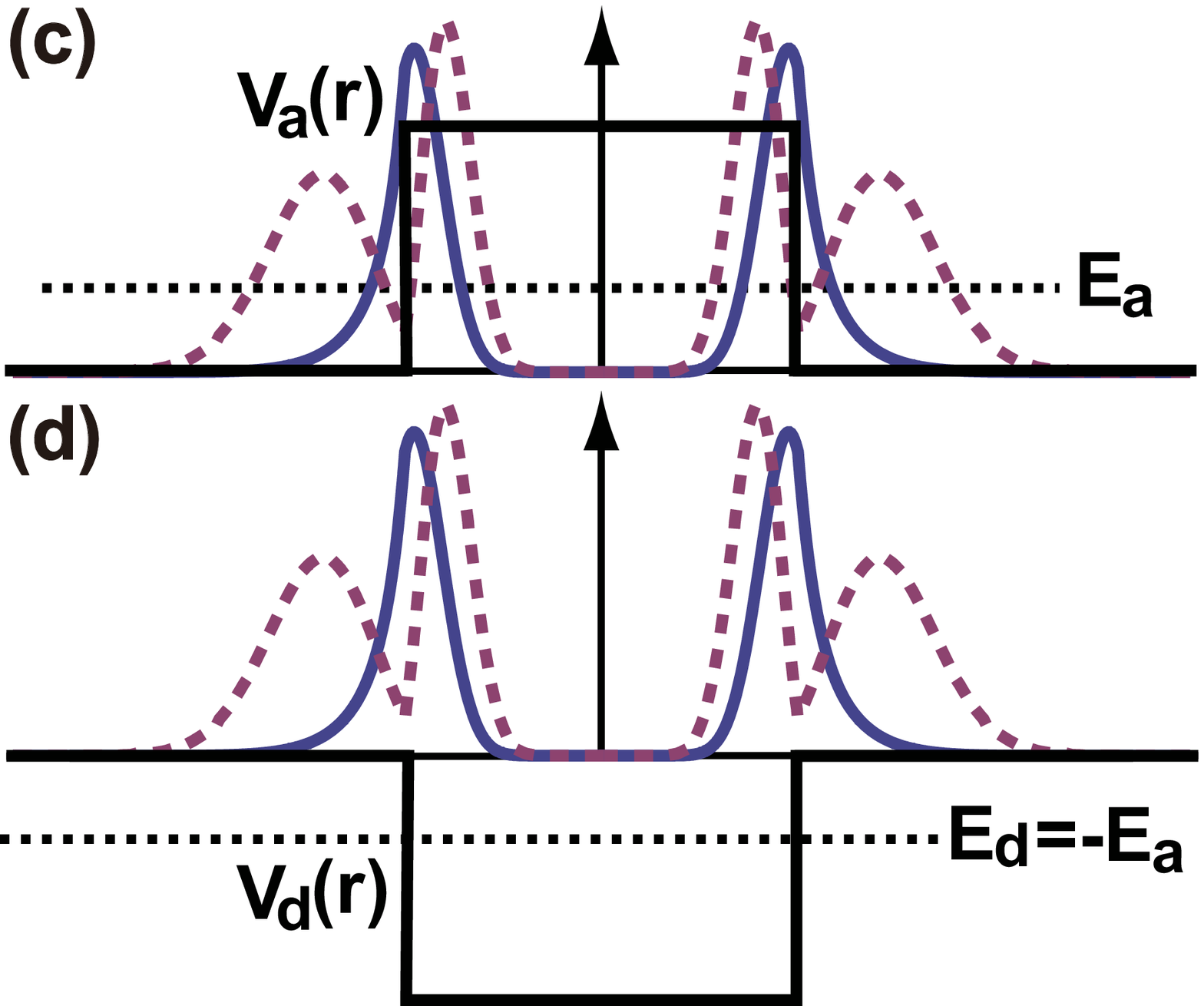}
\includegraphics[width=0.22\textwidth]{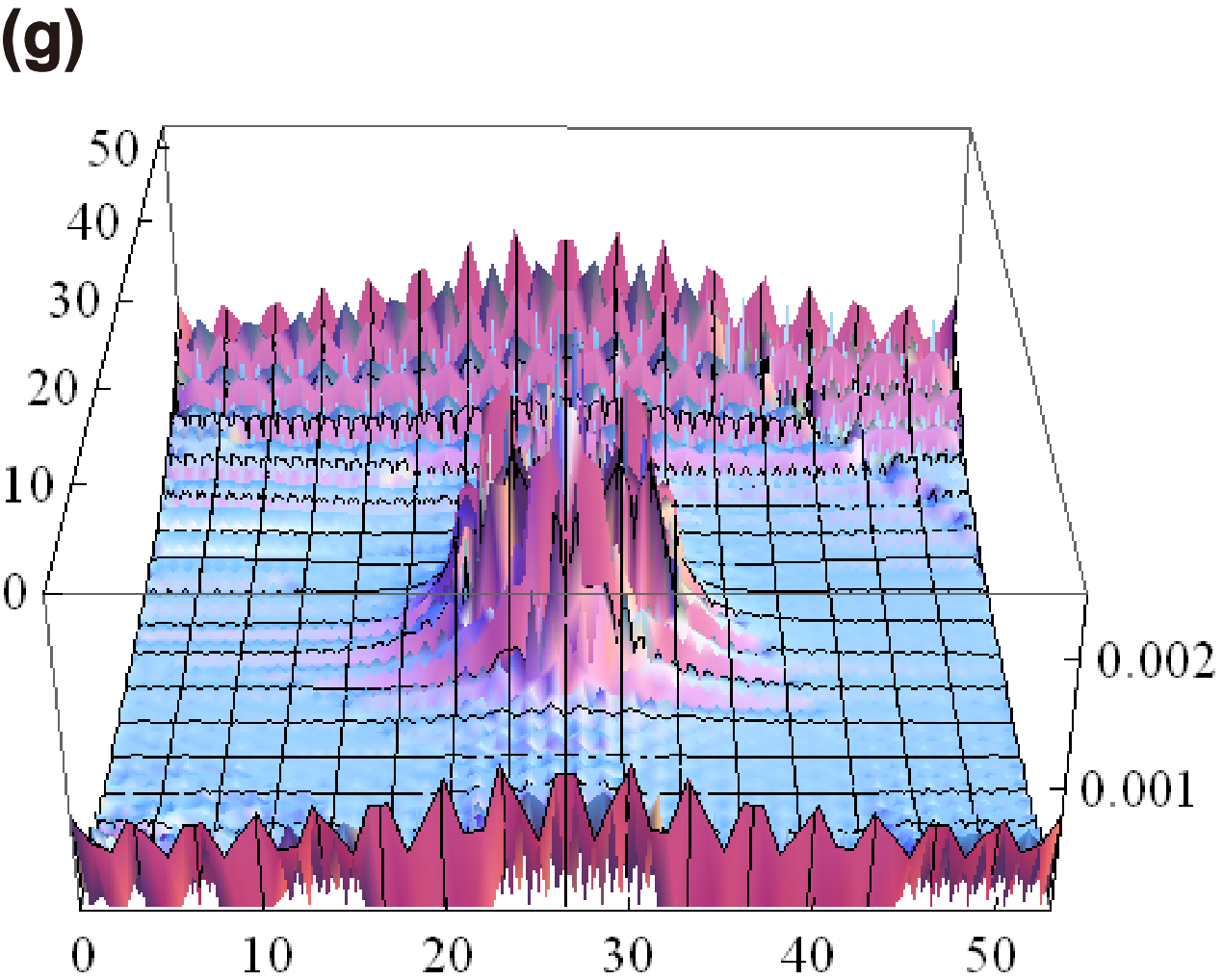}
\caption{Schematic drawings of some typical probability wavefunctions of antidots and dots. We
consider piecewise constant antidot and dot potentials $V_a(r)$ and $V_d(r)$ satisfying
$V_a(r)+V_d(r)=V$. (a) and (b) display dot and antidot states with the same wavefunction but with
different energies $E_d=V-E_a$ and $E_a$.  Solid (dashed) line represents the probability
wavefunction of A (B) component. Naively one may expect that the energy of the antidot state of (a)
should be larger than the dot state by amount $V$ since the antidot state is in the barrier with
the height $V$, but this reasoning is actually incorrect. (c) and (d) are same as in (a) and (b)
but  with $V_a(r)+V_d(r)$=0. Antidot and dot states have energies are  $E_a$ and $E_d=-E_a$. (e)
and (f) are same as  (a) and (b) but wavefunctions are localized outside the antidot and dot
regions. (g) shows tight-binding probability wavefunction of an antidot state of a zigzag
nanoribbon.}\label{fig:antidot1}
\end{center}
\end{figure}

Electron wavefunctions of a graphene quantum dot have been investigated actively both with and
without a magnetic field\cite{Mat,Hew,Sch,Rec,Ando2,Che,Kat,Sta,Sil,Chen,Gia}: In absence of a magnetic field an electron
cannot be localized due to Klein tunneling and only quasibound states are
allowed. However, a perpendicular magnetic field enhances  the localization of the wavefunctions
and both bound states and quasibound states occur.

Recently properties of  antidot lattices\cite{Ped,Van,Shen,Erom,Rosa,Zheng,Peter2}
have  been explored actively. But properties of a single antidot in a magnetic field has not been
studied thoroughly yet. We believe that an antidot created by cutting a hole out of the graphene
sheet\cite{Erom,Ped} and an antidot induced by an electrostatic gate can have different properties.
In an antidot created  by holes skipping orbits that encircle an integer number of elementary
quantum flux defined by the antidot are allowed\cite{Yoshi}: $BA\sim m \phi_0,$ where $A$ is the
enclosing area of the antidot, $m$ is the orbital quantum number and $\phi_0$ is the magnetic flux
quantum. However, some graphene states of  a gate induced antidot potential  may not  obey this
quantization rule since Klein tunneling allows significant penetration into the barrier.  Hence,
graphene antidots may not always support skipping orbits at the edge of an antidot. Also we expect
an antidot defined by a piecewise constant potential may have  several interesting properties in
the presence of a  magnetic field.  This is because of the following feature  of the Dirac
equation: eigenstate wavefunctions of the Dirac equation are also eigenstates of
Schr\"{o}dinger-like equation, where  constant potentials $V$ and eigenvalue $E$ appear together in
the  effective
 energy as $E_{\textrm{eff}}^2=(E-V)^2$.
This is an interesting feature of Dirac electrons but its consequences have not been
fully explored in a magnetic field.
Nature of antidot states of a nanoribbon with zigzag edges may also
be interesting. A nanoribbon has surface edges with various
localization lengths\cite{Fujita,Cas}. It is unclear whether these
chiral zigzag edge states may be coupled to confined non-chiral
antidot states.

In this paper we report on  our investigation of these issues. We consider cylindrically symmetric
and piecewise constant antidot potentials. We find that for each electron state  of a graphene
antidot there is a dot  state  with exactly the same wavefunction but with a different energy, see
Figs.\ref{fig:antidot1}(a)-(f). This  symmetry is a consequence of the appearance of $E_{\textrm{eff}}^2$ in
the Schr\"{o}dinger-like equation. We find that the eigenstates may be divided into three classes:
those that are localized inside the antidot(Fig.\ref{fig:antidot1}(a)), those that have significant weights at the
boundary(Fig.\ref{fig:antidot1}(c)),  and those that are  localized
outside(Fig.\ref{fig:antidot1}(e)). Interestingly we find that probability of finding the electron
in the classically forbidden region of the {\it barrier} can be almost one, as shown in
Figs.\ref{fig:antidot1}(a) and (f). It is a counter example to the usual expectation that the
probability wavefunction is larger in a potential well than in a barrier. This is a consequence of
an interplay between strong Klein tunneling and localization of Landau level wavefunctions. In
zigzag nanoribbons we find strong coupling between some antidot states and zigzag edge states, see
Fig.\ref{fig:antidot1}(g). This effect is unique to antidots of zigzag nanoribbons. Experimental
tests of these effects are proposed.

\section{Symmetry between antidot and dot states:}

We consider a cylindrically symmetric and piecewise
constant potential of an antidot or dot: where the radius of the antidot and dot is R:
$V(r)=V_{\textrm{I}}$ for $r<R$ and $V(r)=V_{\textrm{II}}$ for $r>R$. Around the K point of the
Brillouin zone the Dirac equation of an antidot or dot has the form
\begin{eqnarray}
H=v_F\vec{\sigma}\cdot (\vec{p}+\frac{e}{c}\vec{A})+V(r)\label{eq1}
\end{eqnarray}
with the elementary charge $e>0$, the Fermi velocity  $v_F$,
the Pauli spin matrices $\vec{\sigma}=(\sigma_x,\sigma_y,\sigma_z)$,
and magnetic vector potential
$\vec{A}=\frac{B}{2}(-y,x,0) $. Magnetic field $\vec{B}$ is along z-axis and is perpendicular to
the graphene layer.
Since  the Hamiltonian commutes with the angular momentum operator
$J_{z}=-i\partial_{\varphi}+\sigma_{z}/2$ the eigenstates
can be written  in polar coordinates as
\begin{eqnarray}
\Psi_{jm}(r,\varphi)=e^{i(j-1/2)\varphi}\left(\begin{array}{c}\chi_{A}(r)\\
\chi_{B}(r)e^{i\varphi}\end{array}\right).
\end{eqnarray}
For each angular momentum number  $j=\pm1/2, \pm3/2,\pm 5/2, \cdots$ there are numerous excited
eigenvalues $E_{j,m}$, labeled, in increasing order, by  $m=\cdots, -2, -1, 0, 1,
2,\cdots$($E_{j,m}\geq 0$ for $m\geq 0$ and $E_{j,m}<0$ for $m<0$).

It follows from Eq.(\ref{eq1}) that  A and
B components of the wavefunction $\psi_{\sigma,\alpha}$ with energy $E_a$ satisfy
\begin{eqnarray}
 v_{F}^2\left[\left(\vec{p}+\frac{e}{c}\vec{A}\right)\right]^{2}\psi_{\sigma,\alpha}
= ((E_a-V_{\alpha})^2\mp E_c^2)\psi_{\sigma,\alpha}\label{eq2}
\end{eqnarray}
where the subscript $\alpha=\textrm{I, II}$ stands for $r<R$ or $r>R$ and the index $\sigma=-(+)$
denotes for A (B) component.
 The constants
$E_c=\frac{\hbar v_{F}}{\ell}$ and  $\ell=\sqrt{\frac{\hbar c}{|B|e}}$  stand for the
characteristic energy scale of Dirac electrons in magnetic fields and the magnetic length. If the
potential is not piecewise constant  another term containing $\frac{dV}{dr}$ will be present in
this equation. The equation has the same mathematical structure as the  Schr\"{o}dinger equation of
the two-dimensional Landau levels, but there is an important {\it difference}: the eigenenergy
$E_a$ and the constant potential $V_{\alpha}$ appears together as $E_{\textrm{eff}}^2=(E_a-V_{\alpha})^2$.
For $V(r)=0$ we obtain the usual result of two-dimensional Dirac electrons in a magnetic field,
$E^2=2E_c^2n$. Now consider a cylindrical antidot and dot with piecewise constant potentials
$V_a(r)$  and $V_d(r)$ such that the sum of them is a constant $V_d(r)+V_a(r)=V$. Then, it can be
shown from Eq.(\ref{eq2}) that, when the  antidot potential $V_a(r)$ has an eigenenergy $E_a$,
the dot potential $V_d(r)$ will have an eigenenergy $E_d=V-E_a$ with the {\it same eigenstate}.
Several examples for different pairs of dot and antidot are shown in Fig.\ref{fig:antidot1}.

\begin{figure}[!hbpt]
\begin{center}
\includegraphics[width=0.35\textwidth]{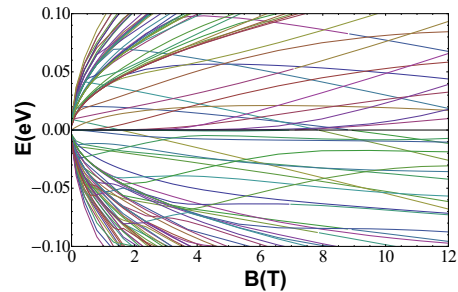}
\caption{Energy  spectrum of antidot as a function of magnetic field. }\label{fig:spectrum}
\end{center}
\end{figure}

\begin{figure}[!hbpt]
\begin{center}
\includegraphics[width=0.3\textwidth]{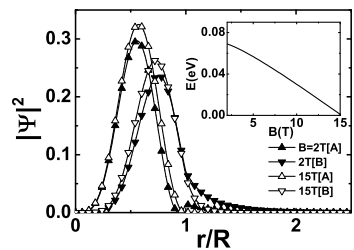}
\caption{ Plot of the  probability wavefunctions of state $j=5/2$ at two different value of $B$.
Inset: its energy level decreases as  $B$
increases. }\label{fig:wave_de}
\end{center}
\end{figure}

\begin{figure}[!hbpt]
\begin{center}
\includegraphics[width=0.15\textwidth]{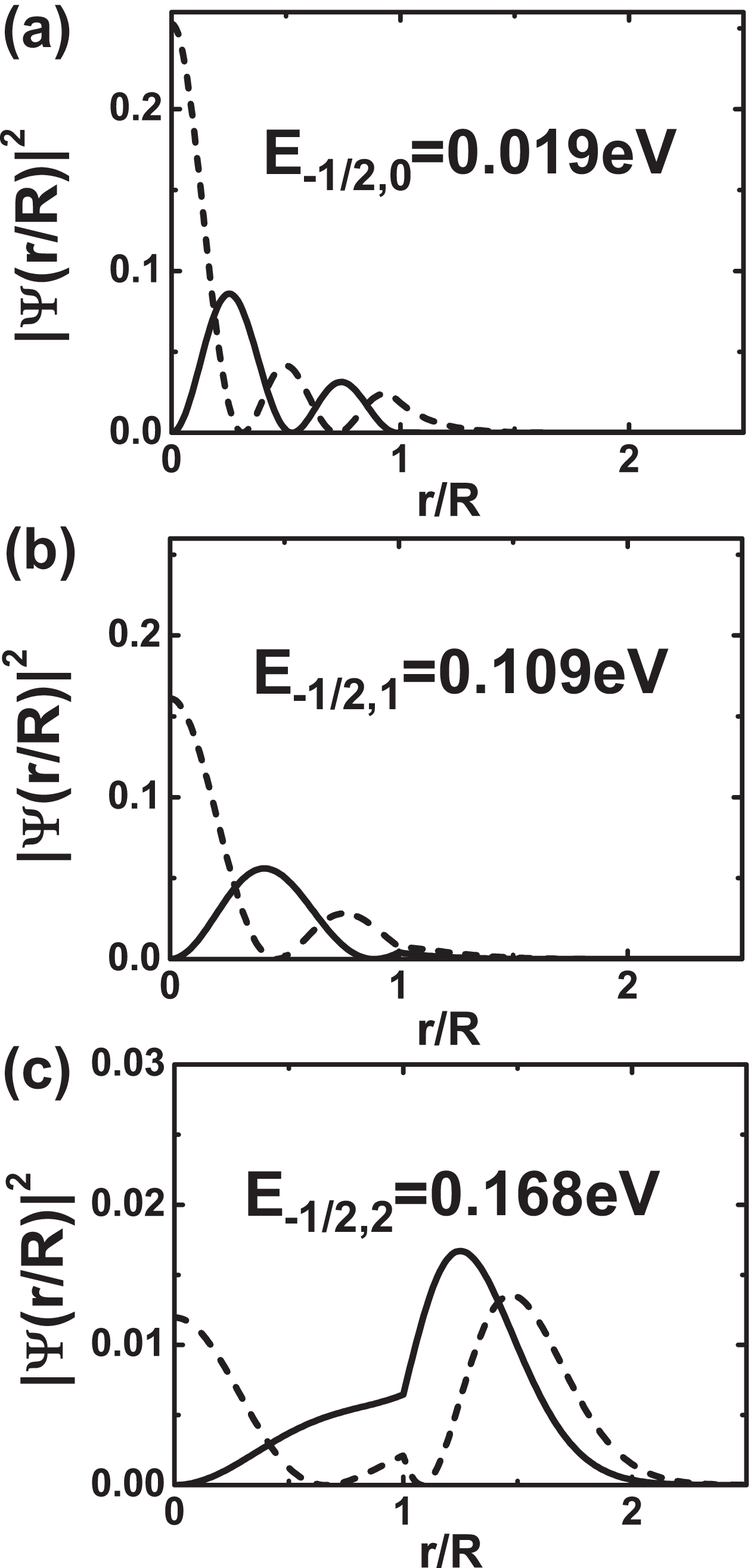}
\includegraphics[width=0.15\textwidth]{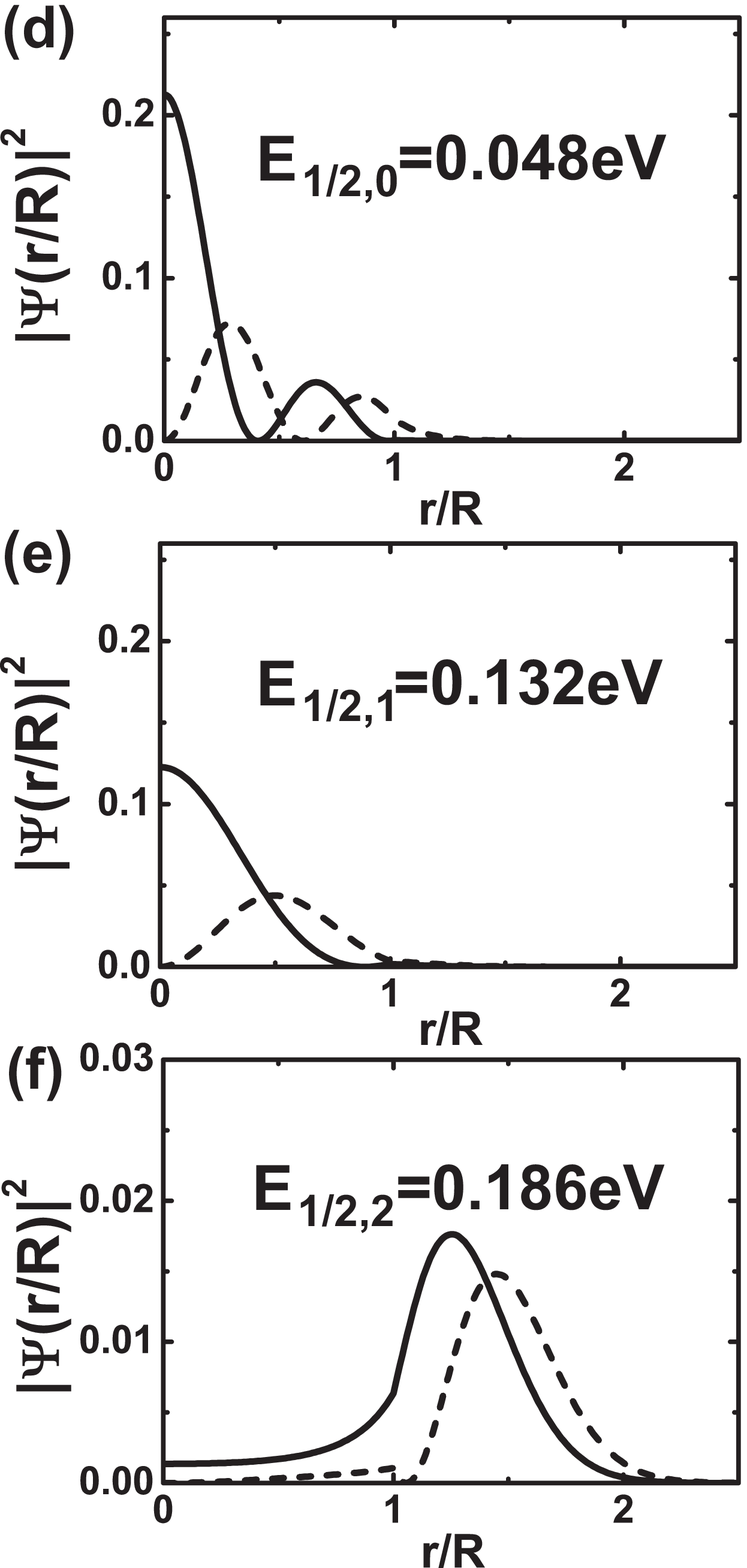}
\caption{Probability wavefunctions of antidot with radius $R$ at $B=10.28T (\ell=80\AA)$, A and B
components are represented by solid and dashed lines.
 }\label{fig:wavefunction1}
\end{center}
\end{figure}

\begin{figure}[!hbpt]
\begin{center}
\includegraphics[width=0.15\textwidth]{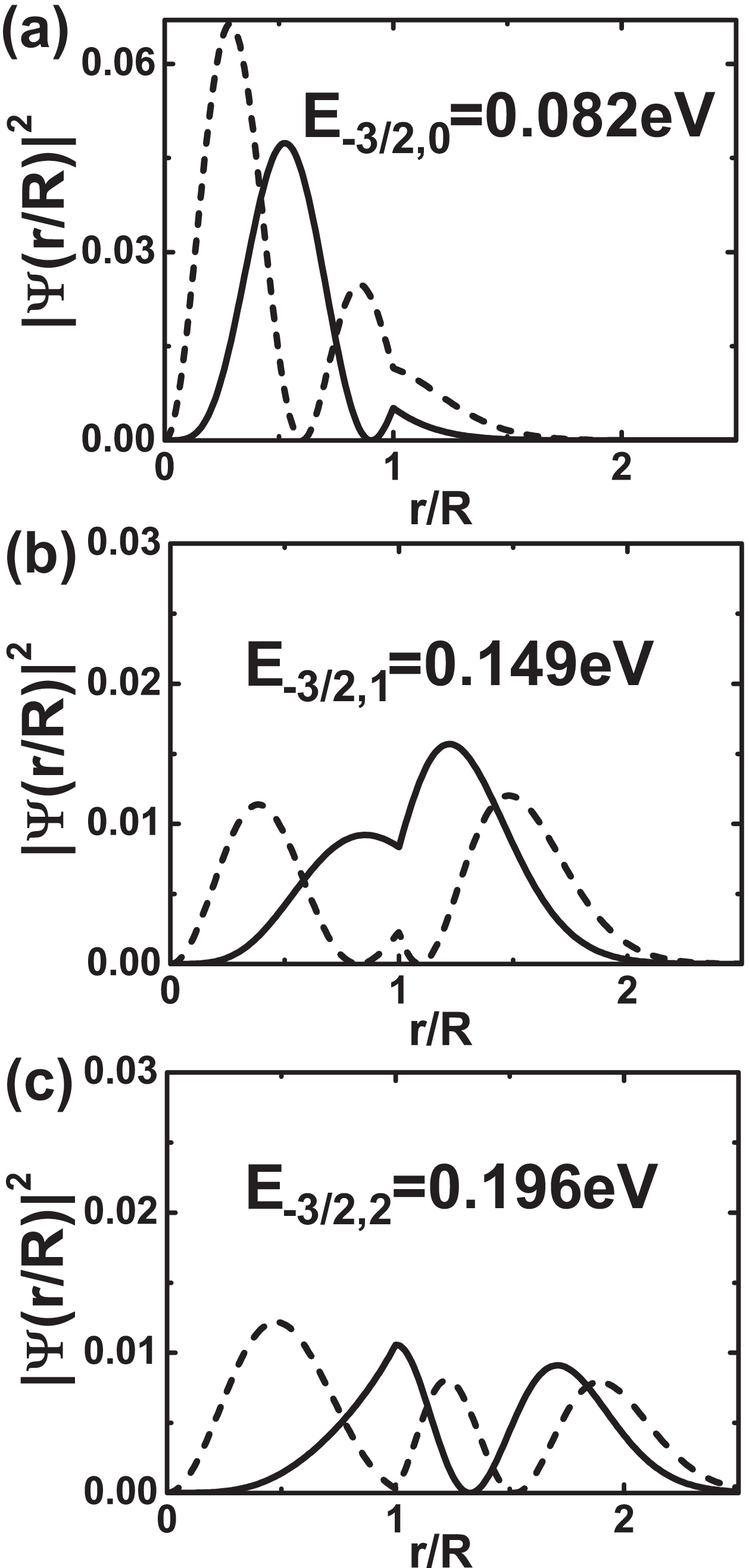}
\includegraphics[width=0.15\textwidth]{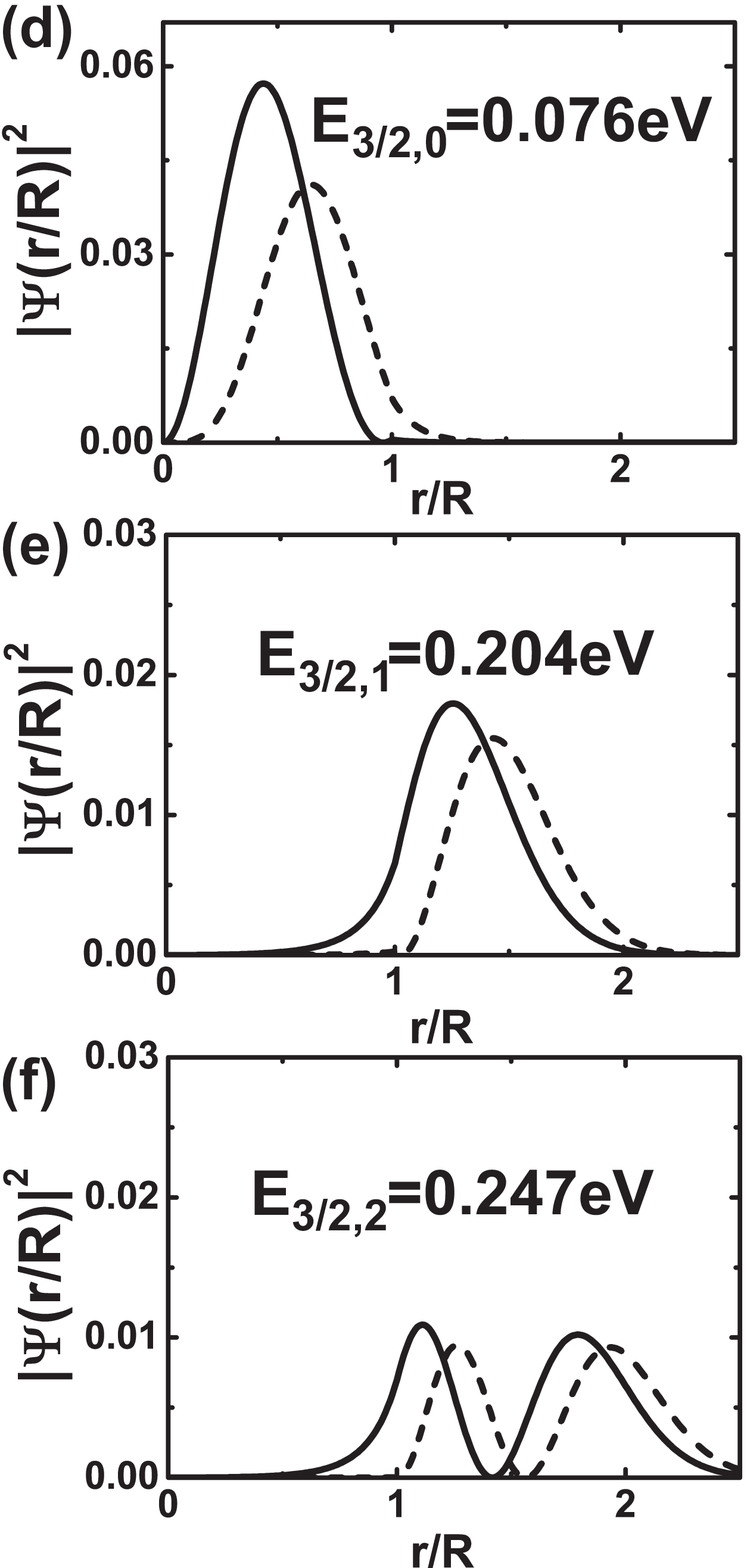}
\caption{Same as in Fig.\ref{fig:wavefunction1}
but for $j=\pm 3/2$.}\label{fig:wavefunction4}
\end{center}
\end{figure}

\section{ Antidot Eigenvalues and eigenstates}

The properties of probability wavefunctions and eigenenergies of a graphene antidot
are rather different from those of the ordinary antidot.  This is because skipping orbits are
not well defined in graphene antidot due to strong Klein tunneling in the presence of a magnetic field.
The eigenstates and eigenvalues of the antidot may be
found by matching solutions inside and outside of the antidot at the boundary $r=R$:  the exact
eigenstate wavefunctions  of Eq.(\ref{eq1}) for  a cylindrical potential can be described by
confluent hypergeometric functions\cite{Sch,Rec} instead of Laguerre polynomials of bulk Landau
levels. The energy spectrum is displayed in Fig.\ref{fig:spectrum} for  $V_{\textrm{I}}=0.26$eV,
$V_{\textrm{II}}=0$, and $R=200\AA$. In the strong magnetic field limit $R/\ell \rightarrow \infty
$ we have tested numerically that the bulk graphene Landau level energies are recovered.

Many  positive (negative) energy levels in Fig.\ref{fig:spectrum} increase (decrease)  as
$\sqrt{B}$. Their wavefunctions are localized outside the antidot and are nearly unaffected by the
antidot potential.  An example is shown in Fig.\ref{fig:antidot1}(e). Other  energy levels display
different dependence on $B$. Their wavefunctions  are affected by the antidot potential and have
significant weights on the boundary $r=R$.
An example is given in Fig.\ref{fig:wave_de}. As one can expect from the usual Landau level physics,
the effective energy $E_{\textrm{eff}}=E-V$ of this level
increases as $B$ increases, and
its wavefunction
becomes more confined in the antidot region.
However, this implies that the eigenenergy $E$ decreases with increasing $B$,
as shown in the inset of Fig.\ref{fig:wave_de}.

We have investigated the eigenfunctions for different values of $E$ and $j$. The first three lowest
positive energy states for several values of $j$ are shown in Figs.\ref{fig:wavefunction1} and
\ref{fig:wavefunction4}. In contrast to ordinary tunneling physics, we observe a significant
penetration into the antidot region: note that for $j=-1/2$ and $-3/2$ the probability
wavefunctions of A component are smaller than probability wavefunctions of B component in the
antidot region. For $j=1/2$ and $3/2$ the opposite is true. The probability wavefunctions of the
lowest energy states with $j=1/2$ are plotted in Fig.\ref{fig:wavefunction1} (d): in the limit
$r\rightarrow 0$ the wavefunction $\chi_{A}(r)$  is proportional to $e^{-\frac{r^{2}}{4\ell^2}}$
while $\chi_{B}(r)$ is proportional to $re^{-\frac{r^{2}}{4\ell^2}}$. However, for $j=-1/2$  the
opposite is true, as shown in Fig.\ref{fig:wavefunction1} (a): $\chi_{A}(r)$ is proportional to
$re^{-\frac{r^{2}}{4\ell^2}}$ and $\chi_{B}(r)$ is proportional to $e^{-\frac{r^{2}}{4\ell^2}}$. We
find that as $|j|$ increases the mean radii of the wavefunctions $\Psi_{jm}$ increase. When the
mean radius $\sqrt{\langle r^2_{jm}\rangle}$ is greater than $R$ the probability wavefunctions will
be peaked in the barrier. In a quantum well the wavefunction for a large $|j|$, for example,
$-51/2$, is thus strongly localized in the {\it barrier}, as shown schematically in
Fig.\ref{fig:antidot1}(f). This unusual effect is possible because the constant potential of the
barrier appears only in $(E-V)^2$(see Eq.(\ref{eq2})). The wavefunction of this state has a rather
different form in comparison  with quasibound states at zero magnetic field with a significant
weight inside the dot\cite{Sil,Gia,Mat}.

\begin{figure}[!hbpt]
\begin{center}
\includegraphics[width=0.2\textwidth]{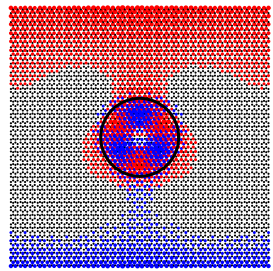}
\caption{ Occupation probabilities on A and B carbon atoms of a state near the Dirac point with
$E=-0.01$eV. Small black dots indicate lattice sites and red (blue) dots indicate values of
probability wavefunction on A (B) carbon atoms.  The circle represents the antidot. The unit of x
and y axis is $a=2.46{\AA}.$ }\label{fig:wave}
\end{center}
\end{figure}

\begin{figure}[!hbpt]
\begin{center}
\includegraphics[width=0.3\textwidth]{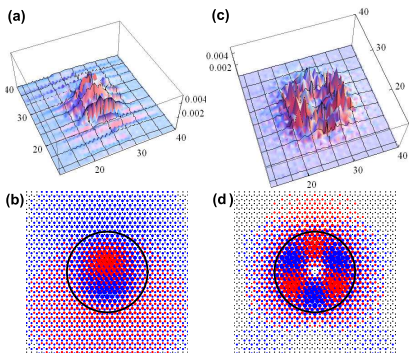}
\caption{(a) and (b) display probability wavefunction of a state with  $E=0.18$ eV and its pseudospin picture,
(c) and (d) show probability wavefunction of a state with  $E=-0.04$ eV and its pseudospin picture.
}\label{fig:wave2}
\end{center}
\end{figure}

\section{Antidot states of  zigzag nanoribbons}

In zigzag nanoribbons it would be interesting to investigate the interplay between Klein tunneling
and backscattering\cite{Ando3,Waka,Waka2}.   We adopt the following model for a
nanoribbon with an antidot. The zigzag edges are horizontal along the x-axis and periodic boundary
condition is imposed on the left and right edges. The horizontal and vertical lengths of the
nanoribbon are $L_x=130.35\AA$ and $L_y=130.64\AA$. A magnetic field of $B=20$T is applied
perpendicular to the graphene sheet.
 The  radius of the antidot is 19.68$\AA$ and its the potential height  is  $V=1$eV.
We solve for eigenstates and eigenvalues in the usual tight-binding Hamiltonian. Fig.\ref{fig:wave}
displays  the occupation probabilities on A and B carbon atoms of a eigenstate close to the Dirac
point with nearly zero energy (its probability wavefunction is shown in Fig.\ref{fig:antidot1}(g)).
This state is a mixture of zigzag edge and antidot states. Opposite chiralities are found on the
opposite sides of the zigzag edges.  Since occupation is  chiral along the zigzag edges no current
flows along the edges.  This mixed state is thus  not a  magnetic edge state.
In the absence of an antidot some states near the Dirac point can have long localization lengths while
states away from the  Dirac point do not\cite{Fujita,Cas}.
Some of these states with localization lengths comparable to the width of the nanoribbon
couple with the antidot, and  significant probability wavefunction can be present both at the edges
and near the antidot.

On the other hand, probability wavefunctions of eigenstates away from the Dirac
point are not peaked along the edges.
Their properties are displayed in Fig.\ref{fig:wave2}.
Figs.\ref{fig:wave2}(a) and (b) show an antidot state with the probability wavefunctions that are
peaked at the origin of the antidot. Note that a similar feature  is observed with Dirac electrons, as shown in
Figs.\ref{fig:wavefunction1}(a) and (d). In contrast, Figs.\ref{fig:wave2}(c) and (d) display a state
with the probability wavefunction that is  zero at the origin, consistent with the corresponding result of Dirac
electrons, see Figs.\ref{fig:wavefunction4}(a) and (d). Tight-binding and Dirac electron models thus
yield qualitatively similar results.

\section{Discussions and conclusions}

\begin{figure}[!hbpt]
\begin{center}
\includegraphics[width=0.2\textwidth]{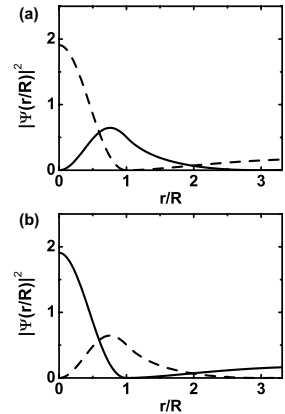}
\caption{ (a) and (b) display probability wavefunctions of a  degenerate pair of K  and K$^\prime$
with $E=0.188$eV. A component is dashed line and  B component is solid line,  These probability
wavefunctions are obtained solving Dirac equations. }\label{fig:dirac}
\end{center}
\end{figure}
\begin{figure}[!hbpt]
\begin{center}
\includegraphics[width=0.25\textwidth]{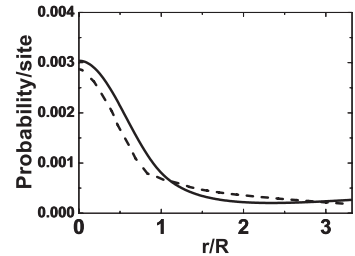}
\caption{
The solid line shows the total probability wavefunction at lattice sites  with
$E=0.188$eV obtained by solving Dirac equation and the dashed line displays the total probability wavefunction
with $E=0.175$eV  obtained by solving the tight-binding Hamiltonian. }\label{fig:compare}
\end{center}
\end{figure}

\begin{figure}[!hbpt]
\begin{center}
\includegraphics[width=0.2\textwidth]{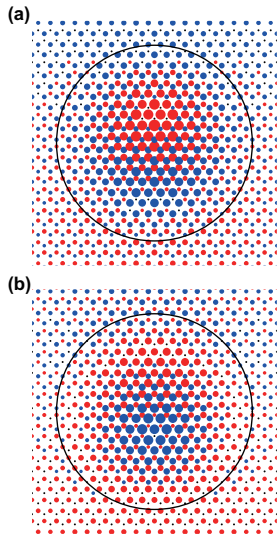}
\caption{ Occupation probabilities on A and B carbon atoms of a state near the Dirac point with
E=0.181eV(a), E=0.175eV(b). The circle represents the antidot. The unit of x and y axis is
$a=2.46$\AA}\label{fig:tight}
\end{center}
\end{figure}

In out treatment of Dirac equations we ignored the coupling of  states of K and K$^\prime$ points
due to the presence of a sharp boundary\cite{Ryc} in the antidot potential. In the absence of the
sharp boundary states of  K and K$^\prime$ points are degenerate with wavefunctions of A and B
components exchanged, as shown in Fig.\ref{fig:dirac} (The radius of the antidot is 19.68$\AA$ and
its potential height is $V=1$eV with the magnetic field $B=20$T). The coupling between  K and
K$^\prime$ points can be included in a tight-binding model, and our calculation shows that this
degeneracy of  Dirac equations is broken slightly: for example, $E=0.188$eV splits into $E=0.175$eV
and $0.181$eV, Moreover, the wavefunctions of tight-binding and Dirac equation are rather close to
each other, see Fig.\ref{fig:compare}. Furthermore, inside the antidot A and B components of nearly
degenerate tight-binding wavefunctions are approximately similar to those of degenerate solutions
of Dirac equations, see  Figs.\ref{fig:dirac} and \ref{fig:tight}: A components are dominant in
Figs.\ref{fig:dirac}(a) and \ref{fig:tight}(a) while B components are dominant in
Figs.\ref{fig:dirac}(b) and \ref{fig:tight}(b). These tight-binding results demonstrate that the
neglect of the coupling between K and K$^\prime$ points in the approach of Dirac equations is a
reasonable  approximation.

In this paper we have discussed several unusual properties of tunneling physics
of step-like graphene potential barriers in magnetic fields. Scanning tunneling
microscope\cite{Niimi} may be used to test strong Klein tunneling effects in antidots since the
electron density in the antidot should be non-zero in contrast to an ordinary antidot. Infrared
optical transitions\cite{Jiang} may be used to test the symmetry between antidot and dot
wavefunctions discussed in this paper: Consider an occupied level $E_{a,1}$ and an unoccupied level
$ E_{a,2}$ in an antidot. Then, according to the proposed symmetry, the absorption transition
$E_{a,1}\rightarrow E_{a,2}$ of the antidot will have the same transition probability as the
absorption transition $E_{d,2}=V-E_{a,2}\rightarrow E_{d,1}=V-E_{a,1}$ of the corresponding dot
system.
We have also found a significant penetration of wavefunctions deep into
the barrier region, as shown in Fig.1(f).  This effect is
absent in
the usual quantum Hall physics in Hall bar geometry, and it
may be worthwhile to explore how it may affect
quantum Hall edges of graphene.

\begin{acknowledgments}
This work was supported by the Korea Research Foundation Grant funded by the Korean Government
(KRF-2009-0074470). In addition this work was supported by the Second Brain Korea 21 Project.
\end{acknowledgments}

\end{document}